# Designing Core Layer in Campus Network Using Software-Defined Networking

**Iwan Setiawan, Azis Wisnu Widhi Nugraha, and Sauqi Asrorul Zaza**

Department of Electrical Engineering, Universitas Jenderal Soedirman, Indonesia

{stwn, azis.wwn}@unsoed.ac.id, sauqiasrorulzaza@gmail.com

**Abstract.** As one of the layers in campus network, core layer or backbone network should provide interconnectivity and routing between internal and external networks, including other campus networks or the Internet. With that requirement, designing this layer to be highly available to interconnect and route traffic is a high priority. By using software-defined networking (SDN) paradigm, we attempt to design core layer in campus network employing best practices of network topology and routing protocol at the layer with actual use case of a core layer in a university's campus network. We use RouteFlow as the SDN platform supporting traditional routing protocol, i.e. Open Shortest Path First (OSPF), over OpenFlow (OF) network infrastructure. The experimental testbed/environment consists of two virtual machines (VMs). The first VM is used as the SDN/OF data plane with Open vSwitch in Mininet network emulator, and the second one representing SDN control plane comprising RouteFlow with POX controller. We evaluated the design by testing the interconnectivity using ping for the OF switches in the topology and hosts that are connected to the switches. We also tracked the route of the packets by monitoring traffic passed through all network interfaces of the switches using tcpdump. This case was evaluated since we need to make sure that packets are routed with the best path from source to destination using OSPF that was implemented in the virtual network at the top of the SDN platform.

**Keywords:** Core layer in campus network; routing; software-defined networking; RouteFlow

## 1. Introduction

A campus network is a network of interconnected local area networks (LANs) within an organisation or enterprise that spans inter-buildings in limited geographic areas. The common design for this network is hierarchical consisting of core, distribution, and access layers [1]. The core layer or backbone of campus network is used to interconnect internal to external networks, which are generally other campus networks or the Internet. The basic services needed for this network are interconnectivity and routing. That makes core layer essential to provide those services with availability in mind.

In modern computer networks, there is a new paradigm for making these networks programmable, flexible, and manageable. The term is software-defined networking (SDN) that mainly separates control from data planes, and put the former plane into an external entity, i.e. SDN server or controller. In traditional networks, these planes are coupled in a network device. According to [2] there are more points to define SDN. These are forwarding decisions are flow-based instead of destination-based, and the network is programmable through applications running on the top of SDN controller.

There are several solutions for making SDN available for the masses, network operators for campus networks in particular. These solutions could be developed purely SDN, bottom-up for all layers, or we could use platforms providing frameworks to ease SDN implementation with specific use cases. RouteFlow [3] is one the SDN platforms that provides a way to use traditional Internet Protocol (IP) routing over SDN/OpenFlow infrastructure. It is not just giving network operator opportunities to experiment with SDN infrastructure with traditional IP routing stacks on the top of it, but also a mean to connect to legacy networks and also to ease the migration to SDN.

*Related Works*. Rizvi et al. [4] implemented and evaluated SDN based on RouteFlow in Internet Service Provider (ISP) networks using case study on multiple tests. Sudiyatmoko, Hertiana, and Negara [5] have done a research using RouteFlow with Intermediate System-to-Intermediate System (IS-IS) routing protocol evaluating Quality of Service (QoS) and controller performance for several network topologies. An SDN research using the platform also conducted by Negara and Tulloh [6] experimenting OSPF protocol implemented over the platform with QoS and convergence time parameters for different topologies. They did not designed their research to relate to campus network specifically core layer or backbone with its requirements such as network topology and routing protocol, also its actual use case in a real campus network. Moreover, the evaluation of how the packets were route from source to destination according to the routing protocols that were implemented on the top of SDN/RouteFlow were not carried out by the authors.

*Contributions*. Our contribution in this paper is as follows. We design a core layer in campus network using SDN/OpenFlow based on RouteFlow with triangle for the network topology and OSPF protocol for the routing service. We consider the design with actual topology of a core layer in campus network in Universitas Jenderal Soedirman (Unsoed). Further, we conducted interconnectivity and packet routing tests to evaluate the proposed design.

## 2. Research Method

In this section, we describe the characteristics of the core layer in campus network and put them into requirements that are needed for the design of the layer using SDN/OpenFlow. After that, we propose the design in the form of network topology in a simple architecture of core layer and setup an experimental virtual testbed or environment. Ultimately, we run two tests to validate the design with interconnectivity and packet routing aspects.

### *2.1. System Design*

#### *2.1.1. Network Topology*

There are two choices for choosing network topology for the core layer, square or triangle topologies. It is considered optimum to use triangle topology, rather than the square one for the layer, since it gives equal-cost redundant paths for the best deterministic convergence [1]. In addition to that, point-to-point links between network devices in the topology are desired to take the benefits of quick propagation when changes happened in terms of link up or down.

Unsoed has a campus network with hierarchical network design [7] including a core layer utilising triangle topology connecting three routers which are resided in three different locations: Duren Tiga, Grendeng, and Blater with link bandwidth of 1 Gbps. Figure 1 shows the hierarchical network design in the campus network with the core layer at the top of the architecture. This research where also meant to be an experiment to use SDN with the case study of the core layer at the university, in the hope that the results would give valuable information regarding architecture, requirements, and issues regarding implementation of traditional routing protocol especially OSPF in triangle topology on the top of SDN/OpenFlow infrastructure.

#### *2.1.2. Routing Protocol*

After the network topology is formed using triangle, the network devices will be connected to each other via the physical topology. Routing service on the network devices should compute and decide how packets will travel according the best path from source to destination. This service is usually applied through static or dynamic routing. In the case of core layer, there are various IP routing protocols that can be used in the layer supporting dynamic routing, such as RIP, IS-IS, OSPF. The latter protocol is considered eminent compared to the others. It is open and widely use protocol that has been used in internal network or intra-Autonomous System (intra-AS) routing [8]. This link-state protocol implements Dijkstra's least-cost or shortest-path (SPF) algorithm. It is one of the protocols that has been standardized by Internet Engineering Task Force (IETF) [9]. The core layer in Unsoed's campus network also uses this protocol to dynamically route packets in the network.

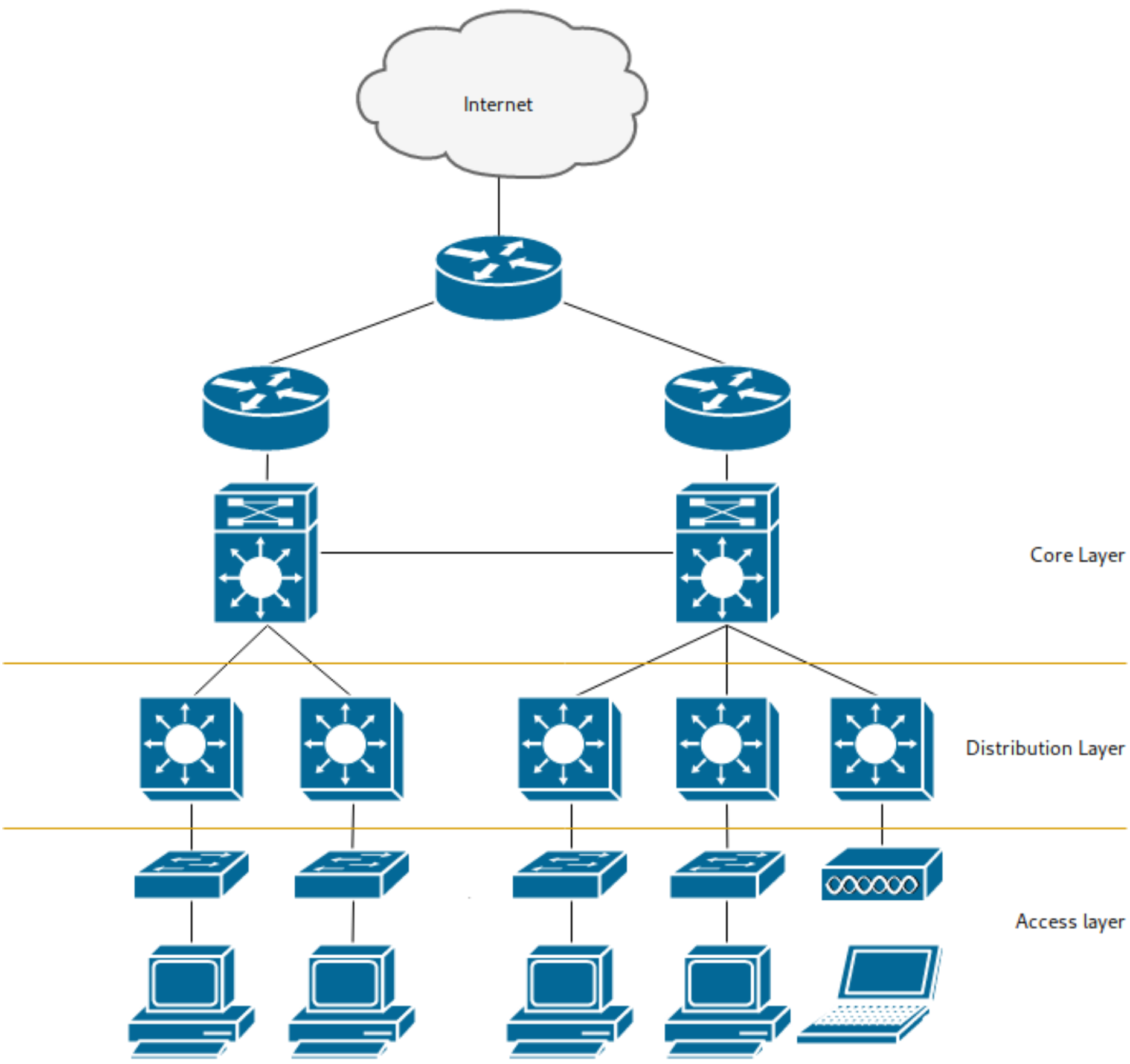

Figure 1. Hierarchical network design in a campus network.

One of the challenges in designing a core layer using SDN is implementing traditional IP over SDN/OpenFlow infrastructure. Our goal of this research is how to apply traditional IP routing, i.e. OSPF, in core layer over SDN/OpenFlow infrastructure.

*2.1.3. SDN for Core Layer*

OpenFlow (OF) [10] that is considered the origin of SDN, caters specifications needed for the communications between controller to switches or vise versa. It is called southbound communications with the use of Application Programming Interface (API). The goal of the project primarily to develop protocol for southbound communications and with that hopefully to seed the innovation through programmability developed at SDN application layer. This project intents to push this in campus networks since there are many opportunities to experiment the new technologies within this type of networks.

RouteFlow [5] is one of the SDN platforms providing virtualised IP routing services over OpenFlow network infrastructure. It is consists of three core components, RFProxy, RFServer, and RFClient along with RF protocol to communicate the two latter elements. This platform uses an SDN/OpenFlow controller, either POX or Ryu. At the top of the platform, there is a virtual network that reproduces the interconnectivity of a physical or virtual infrastructure. The network could run legacy routing engines such as Quagga, BIRD, XORP. Thus, it will support many traditional IP routing protocols supported by those engines, e.g. RIP, IS-IS, OSPF, BGP.

There are several options in choosing SDN controller with different features, code bases, and potentials for future research. In this research, we chose POX as the SDN controller since it is the base controller for experimenting with SDN particularly using Mininet, also it is included in RouteFlow package. Additionally, by choosing this controller, we hope that we just need to explore a small but functional controller program, considering the scope of this research.

Summarising from the aforementioned requirements and considerations for the design, we put network components and preferences to Table 1. The design of the core layer using SDN/OpenFlow is shown in Figure 2, it is adapted from the traditional one on the left. RouteFlow with POX controller will be resided in the SDN Controller. With this information, we designed the experimental network through setup of softwares utilising virtualisation and other supporting ones. This will be described in the next section of the article.

Table 1. Network components and preferences.

| Network Component | Preference |
|---|---|
| Network topology | Triangle with 1 Gbps links |
| Routing protocol | OSPF |
| SDN | RouteFlow with POX controller |

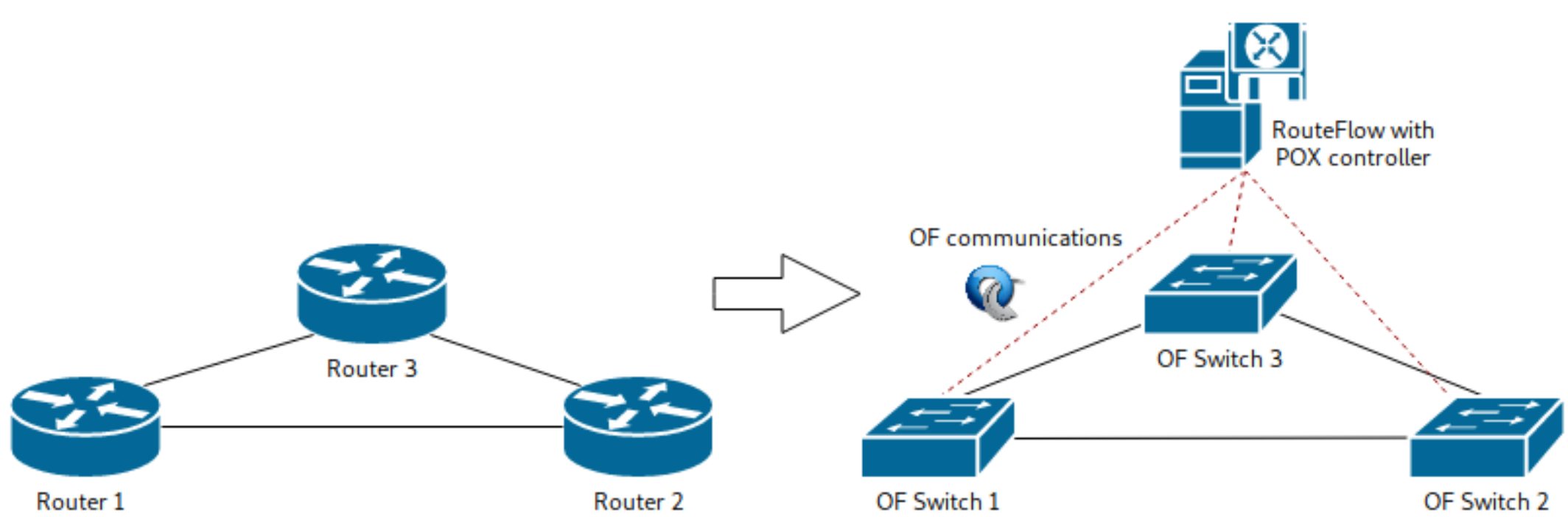

Figure 2. A design of core layer using SDN/OpenFlow converted from the traditional one.

### 2.2. *Experimental Setup*

This experimental research uses a virtual testbed/environment consisting of two virtual machines (VMs) that run on a host, a laptop with specifications of Intel Core i5-6200U with 2 cores at base clock 2.3 GHz and main memory of 12GB. This laptop was installed with Debian GNU/Linux 9 (stretch) 64 bit.

Table 2. Softwares and their functions for this research.

| Software | Ver. | Function | Notes |
|---|---|---|---|
| QEMU/KVM | 2.8.1 | Virtualisation | For creating VMs, installed on host |
| Dnsmasq | 2.76 | DHCP server | Providing network config. for VMs, installed on host |
| Wireshark | 2.2.6 | Network analysis | For analysing data from tcpdump, installed on host |
| RouteFlow | CPqD | SDN platform | POX controller and RF*, installed in RouteFlow VM |
| Open vSwitch | 2.0.2 | Data Plane | Included in Mininet VM |
| Tcpdump | 4.9.0 | Network monitoring | Included in Mininet VM |

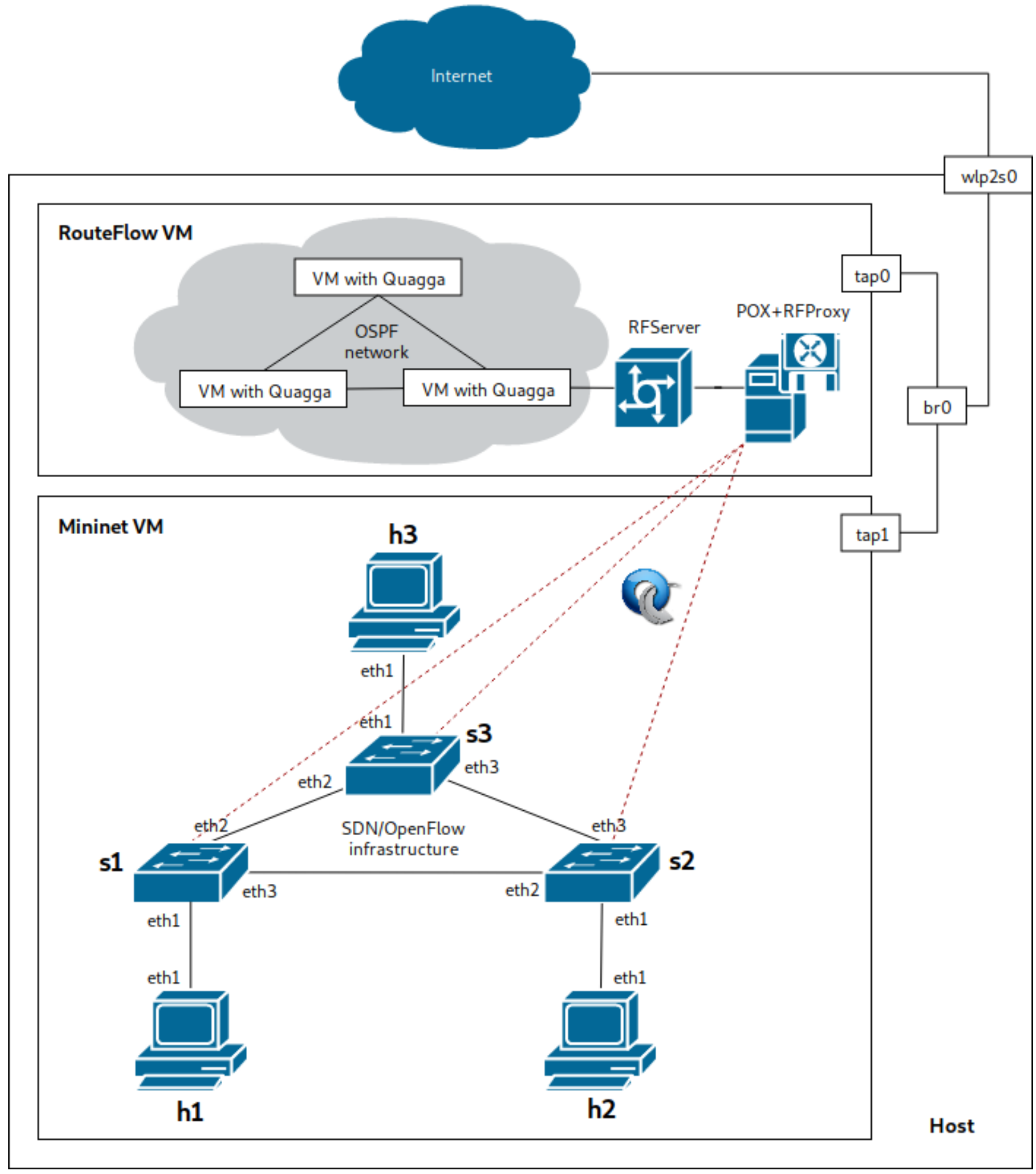

Figure 3. The experimental setup of VMs along with the topology used in this research.

Table 3. Network interface configurations for switches and hosts.

| **Switch/Host** | **Network Interface** | | |
|---|---|---|---|
| | **Ethernet 1 (eth1)** | **Ethernet (eth2)** | **Ethernet (eth3)** |
| Switch 1 (s1) | 172.31.1.1/24 | 10.0.0.1/24 | 20.0.0.1/24 |
| Switch 2 (s2) | 172.31.2.1/24 | 20.0.0.2/24 | 30.0.0.2/24 |
| Switch 3 (s3) | 172.31.3.1/24 | 10.0.0.3/24 | 30.0.0.3/24 |
| Host 1 (h1) | 172.31.1.100/24 | - | - |
| Host 2 (h2) | 172.31.2.100/24 | - | - |
| Host 3 (h3) | 172.31.3.100/24 | - | - |

On the top of the host, we used several softwares for this research that can be shown in Table 2. Internet connection is also needed for setting up the experimental environment particularly when updating packages list and installing packages from repositories, also cloning RouteFlow repository and creating the virtual routers including all dependencies.

We created an experimental setup to accomodate the design and include the softwares stated in Table 2 so that it forms an integrated virtual testbed/environment for this research. This setup can be seen in Figure 3. The network configuration for each network interface in OF switches and hosts is shown in Table 3.

*2.2.1. Host*

Our laptop host was installed with softwares required especially QEMU/KVM and dnsmasq. We installed QEMU/KVM, dnsmasq, and their dependencies, then we created scripts to run two VMs with hardware-supported virtualisation for RouteFlow and Mininet VMs on host. The specifications of the VMs can be described in Table 4.

It was also required to configure two TUN interfaces (virtual network interfaces) and one virtual bridge so that the two VMs connected to the host machine via a virtual network. Option such as "macaddr" was included for making each VM has different MAC addreses for its network interface and did not have the same IP address. Internet connection was provided using NAT set on host.

Table 4. Specifications for the VMs used in this research.

| **Specifications** | **RouteFlow VM** | **Mininet VM** |
|---|---|---|
| Virtual environment | Dual-core, RAM 4GB, Disk 20GB | Dual-core, RAM 3GB, Disk 20GB |
| Operating System | Ubuntu 12.04.5 64 bit | Ubuntu 14.04.5 64 bit |
| Function | RouteFlow using POX controller | Data Plane using OVS in Mininet |

*2.2.2. RouteFlow*

RouteFlow as the SDN platform we used in this experiment should be setup in certain steps. We used CPqD's version of RouteFlow [11] since it is stable even though rather old than the community one. In general, the steps are preparing the system inside the RouteFlow VM including installing packages needed for compiling and running RouteFlow, downloading source code of CPqD's RouteFlow, compiling RFClient, configuring three virtual routers for Quagga as well as OSPF forming triangle topology, and finally running a script to generate LXCs for the virtual routers. The configuration for the latter includes a csv file defining how many LXCs are built, VM ID, data path ID, and port density.

In each of the virtual router's LXC configuration, we set port density, MAC address, interface name, and link via virtual bridge. We have set the OSPF configuration to 1 second and 4 second for hello and dead intervals respectively plus configured networks for backbone area. Regarding the configuration of Quagga, we defined interfaces and IP addresses for each of the virtual router. After all steps are run successfully, we execute a script to run RouteFlow with POX controller including all softwares needed.

*2.2.3. Mininet*

Mininet [12] is a network emulator for experimenting SDN on a computer or laptop. This emulator includes controllers and data plane such as Open vSwitch (OVS). We use this emulator to simulate SDN data plane using OpenFlow switches based on Open vSwitch.

Mininet provides VM image to be downloaded from the official repository. We used version 2.2.2 and OpenFlow 1.0 for this research. There is no installation of packages in the Mininet VM since it provides all packages needed. The details of the softwares which are included can be seen in Table 2.

Within the Mininet as the data plane of our experiment, we need to create triangle topology using Mininet API. Thus, we wrote a Python script consisting all needed configuration for creating this topology. The script will be run as a parameter of Mininet binary along with passing other parameters such as defining custom topology, remote controller to RouteFlow VM, and link bandwidth set to 1 Gbps. An additional file is needed to configure default route of networks where each host is connected.

## 3. Evaluation and Results

We run the evaluation of the design according to the method aforementioned. Our aims for this evaluation are for testing the interconnectivity of network switches as well as hosts that are connected to each switch, also route tracing of packets from source to destination. We use ping program to send Internet Control Message Protocol (ICMP) packets from source to destination hosts and measure the round-trip time (RTT) for the interconnectivity and packets captured in all interfaces of the switches to obtain the routes.

### *3.1. Interconnectivity*

Evaluation of the interconnection between network devices also a host connected to each device needs to be tested. As aforestated, we use ping using combination of host pairs 10 times. The combination are host 1 to host 2 (h1 to h2), host 2 to host 3 (h2 to h3), and host 3 to host 1 (h3 to h1). We also check the captured data from tcpdump to get the information of OpenFlow packet-ins and -outs that encapsulate ICMP packets that pass through the switches. Table 5 shows the results of the interconnectivity test.

Table 5. Results of interconnectivity evaluation of the design.

| Test | Link Test | | | | | |
|---|---|---|---|---|---|---|
| | h1 to h2 | | h2 to h3 | | h3 to h1 | |
| | RTT | OF Packet | RTT | OF Packet | RTT | OF Packet |
| 1 | 31.5 | Yes | 46.4 | Yes | 39.4 | Yes |
| 2 | 71.8 | Yes | 25.8 | Yes | 54.2 | Yes |
| 3 | 31.9 | Yes | 28.1 | Yes | 46.8 | Yes |
| 4 | 72.4 | Yes | 42.5 | Yes | 55.0 | Yes |
| 5 | 62.4 | Yes | 18.7 | Yes | 26.9 | Yes |
| 6 | 38.4 | Yes | 41.8 | Yes | 154 | Yes |
| 7 | 34.4 | Yes | 12.2 | Yes | 40.9 | Yes |
| 8 | 49.4 | Yes | 34.4 | Yes | 54.7 | Yes |
| 9 | 71.9 | Yes | 45.3 | Yes | 126 | Yes |
| 10 | 33.7 | Yes | 110 | Yes | 51.7 | Yes |
| Avg. | 49.841 | | 40.644 | | 65.129 | |

We can see on Table 5 that the range of the average values is 40-65 ms. These results are considered normal, but if we consider the RTT values for LANs these are bigger.

### *3.2. Packet Routing*

The second evaluation of the design is packet routing. We need to make sure that ICMP packets are sent from source to destination following the shortest/optimum path according to the OSPF protocol.

Tcpdump was setup to monitor all of the interfaces on the switches. This can be seen on Figure 4. We use tcpdump since it is light, does not consume more resources, and is included in the Mininet VM. To start with the packet capturing, we need to remote access Mininet VM and start sessions for all switches and run tcpdump instances on them. Since the size of the storage inside VMs was limited, we transfered data generated by tcpdump to host. After that, we could analyse the data using Wireshark. Table 6 shows the results of packet route evaluation in this research.

In the first test, we pinged from h1 to h2. It is shown that the packets were sent via switch 1 (s1) dan switch 2 (s2) since this path is the shortest/best to connect the two hosts. The results are the same with ping tests from h2 to h3 and h3 to h1, packets were routed to the shortest path. The path was computed in virtual routers at the top of the RouteFlow platform using OSPF and further applied the routing information into the SDN/OpenFlow data plane/infrastructure in Mininet via RFProxy.

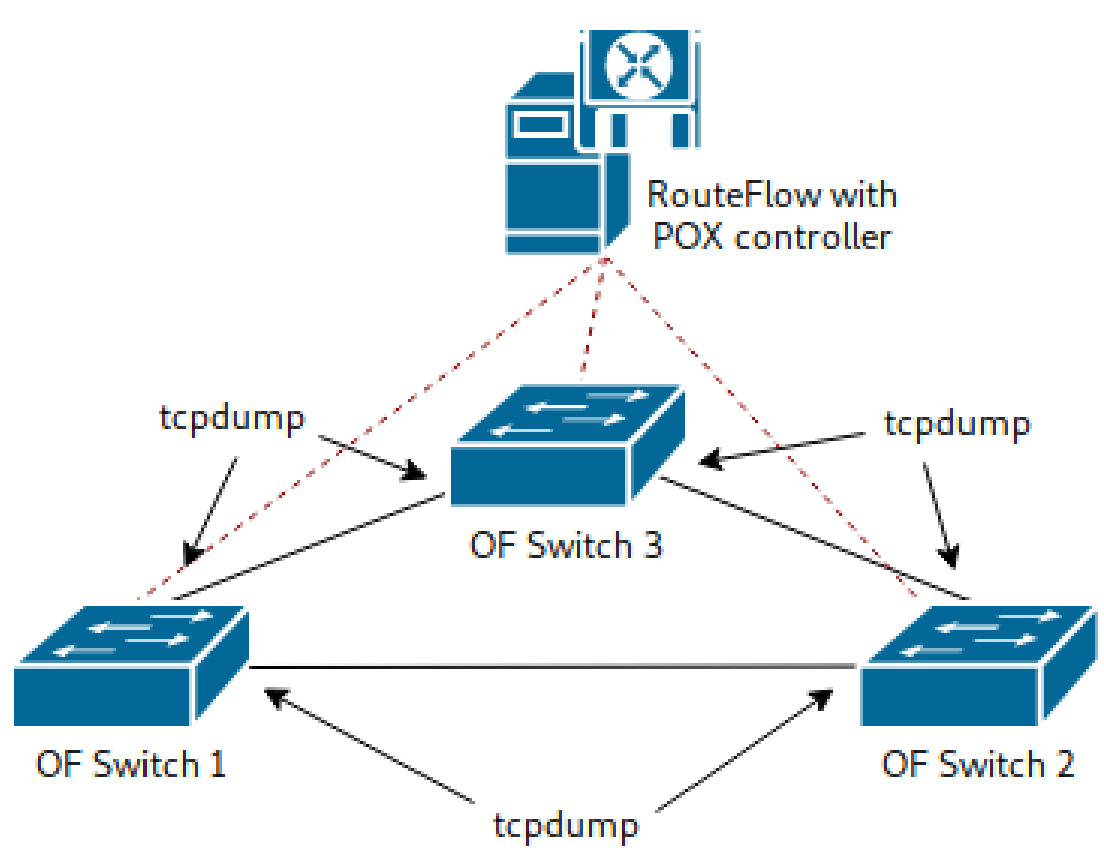

Figure 4. Packet capturing with tcpdump on all switches' interfaces.

Table 6. Results of packet route evaluation of the design.

| Test | s1 | | | s2 | | | s3 | | |
|---|---|---|---|---|---|---|---|---|---|
| | **s1-eth1** | **s1-eth2** | **s1-eth3** | **s2-eth1** | **s2-eth2** | **s2-eth3** | **s3-eth1** | **s3-eth2** | **s3-eth3** |
| h1-h2 | 1 | 0 | 1 | 1 | 1 | 0 | 0 | 0 | 0 |
| h2-h3 | 0 | 0 | 0 | 1 | 0 | 1 | 1 | 0 | 1 |
| h3-h1 | 1 | 1 | 0 | 0 | 0 | 0 | 1 | 1 | 0 |

## 4. Conclusions and Future Work

We have designed a core layer or backbone in campus network using triangle topology, OSPF protocol, and SDN based on RouteFlow with virtual testbed/environment consisting of two VMs. The first VM running RouteFlow and POX controller plus supporting softwares, and the second one running Mininet with Open vSwitch. The two VMs simulate SDN control and data planes providing experimental testbed for the design.

The design has been tested in terms of interconnectivity using ping with all network switches are interconnected and each host representing different network connected to each of the switches. Pairs of hosts were pinged resulting range of RTT average values of 40-65 ms. Packet routing was checked using tcpdump. The tool collected network traffic for all interfaces on the switches, and then we analysed the data with Wireshark to find the ICMP packets that passed through the interfaces. The results are corresponded with SPF algorithm implemented by OSPF protocol which is packets are routed to the shortest path from source to destination.

We plan to develop scenarios with the same experimental setup to evaluate the design in terms of availability of the interconnectivity service against link failure in each possible link in the topology and measure the RTT plus convergence time between the failure and the converged network providing alternate paths. The results would be compared to the state-of-the-art with consideration of the complexity of the topologies.

The implementation of this experimental research on a physical SDN testbed using OpenFlow switches with commodity hardware, e.g. OpenWrt-flashed low-cost switch, would be interesting to provide more real application of the design. The results of this research would give more valuable insights if we put them together with the results of the design in virtual SDN testbed side by side.

**Acknowledgements**

We would like to thank to the Institute of Information Technology and Systems Development (LPTSI) Unsoed for providing us the details of Unsoed's campus network particularly its core layer through their staff in the Center of Infrastructure Development and Services.

**References**

[1] "Campus Network for High Availability Design Guide," *Cisco*. [Online]. Available: https://www.cisco.com/c/en/us/td/docs/solutions/Enterprise/Campus/HA_campus_DG/hacampusdg.html. [Accessed: 07-Nov-2018].

[2] D. Kreutz, F. M. V. Ramos, P. Esteves Verissimo, C. Esteve Rothenberg, S. Azodolmolky, and S. Uhlig, "Software-Defined Networking: A Comprehensive Survey," *Proceedings of the IEEE*, vol. 103, no. 1, pp. 14–76, Jan. 2015.

[3] RouteFlow, https://routeflow.github.io/RouteFlow/

[4] S. N. Rizvi, D. Raumer, F. Wohlfart, and G. Carle, "Towards carrier grade SDNs," *Computer Networks*, vol. 92, pp. 218–226, Dec. 2015.

[5] A. R. Sudiyatmoko, S. N. Hertiana, and R. M. Negara, "Analisis Performansi Perutingan Link State Menggunakan Algoritma Djikstra Pada Platform Software Defined Network (SDN)," *Jurnal Infotel,* vol. 8, no. 1, pp. 40–46, 2016.

[6] R. M. Negara and R. Tulloh, "Analisis Simulasi Penerapan Algoritma OSPF Menggunakan RouteFlow pada Jaringan Software Defined Network (SDN)," *Jurnal Infotel*, vol. 9, no. 1, 2017.

[7] I. Setiawan, A. W. W. Nugraha, and A. S. Atmaja, "Unjuk Kerja IP PBX Asterisk dan FreeSWITCH pada Topologi Bertingkat di Jaringan Kampus," *Jurnal Infotel*, vol. 9, no. 3, Aug. 2017.

[8] J. F. Kurose and K. W. Ross, *Computer Networking: A top-down approach*, 7th ed., Pearson Education Limited, 2017.

[9] OSPF Version 2, https://tools.ietf.org/html/rfc2328

[10] N. McKeown *et al.*, "OpenFlow: enabling innovation in campus networks," *ACM SIGCOMM Computer Communication Review*, vol. 38, no. 2, p. 69, Mar. 2008.

[11] RouteFlow, https://github.com/CPqD/RouteFlow

[12] B. Lantz, B. Heller, and N. McKeown, "A network in a laptop: rapid prototyping for software-defined networks," in *Proceedings of the 9th ACM SIGCOMM Workshop on Hot Topics in Networks*, p. 19, 2010.